\documentclass[
  reprint,
  aps, prl,
  longbibliography,
  nofootinbib,
  floatfix,
  superscriptaddress
]{revtex4-2}
\usepackage{graphicx}
\usepackage{xcolor}
\usepackage[utf8]{inputenc}
\usepackage[normalem]{ulem}
\usepackage{amsmath, amsfonts, amssymb}
\usepackage[hidelinks]{hyperref}
\usepackage{physics, tabularx}
\usepackage{natbib}
\newcommand{\prlsection}[1]{%
  \par\medskip
  \emph{#1}%
  ---%
}

\begin{document}

\title{Quantum Mpemba effect in holography}
\author{Xian-Hui Ge}\email[E-mail: ]{gexh@shu.edu.cn}
\affiliation{%
Department of Physics, Shanghai University, 99 Shangda Road, 200444 Shanghai, China
}

\author{Shuta Ishigaki}\email[E-mail: ]{shutaishigaki@shu.edu.cn}
\affiliation{%
Department of Physics, Shanghai University, 99 Shangda Road, 200444 Shanghai, China
}

\author{Yu-Qi Lei}\email[E-mail: ]{yuqi\_lei@shu.edu.cn}
\affiliation{%
Department of Physics, Shanghai University, 99 Shangda Road, 200444 Shanghai, China
}
\affiliation{%
Department of Mathematics, Shanghai University, 99 Shangda Road, 200444 Shanghai, China
}

\author{Yu Tian}\email[E-mail: ]{ytian@ucas.ac.cn}
\affiliation{%
School of Physical Sciences, University of Chinese Academy of Sciences,
1 Yanqihu East Road, 101408 Beijing, China
}

\begin{abstract}

We investigate the quantum Mpemba effect in a holographic superfluid, in which states with stronger initial symmetry breaking relax faster toward the symmetry-restored equilibrium.
%We demonstrate its emergence by identifying the energy flux into the black hole horizon as monotonic distance measure.
We demonstrate its emergence by identifying the shifted free energy computed from the energy flux into the black hole horizon as monotonic distance measure.
By decomposing the nonlinear bulk dynamics based on quasinormal modes, we reveal that the anomalous relaxation is governed by a dynamical competition in which the slowest-decaying mode is suppressed while the second mode is amplified.
These findings provide a holographic perspective on the quantum Mpemba effect in nonequilibrium relaxation involving strongly coupled degrees of freedom.
%These results establish that, within the holographic dual description, the QME in a nonlinear quantum system can be understood as a dynamical competition among quasinormal modes.

\end{abstract}

\maketitle

\prlsection{Introduction}
%The Mpemba effect describes a counterintuitive phenomenon where a hotter system cools faster than a colder one~\cite{Mpemba_1969}, which has \textcolor{blue}{transcended} its classical origins to become an important problem in nonequilibrium statistical physics. 
%As a paradigmatic phenomenon in nonequilibrium statistical mechanics,
The classical Mpemba effect is the counterintuitive phenomenon in which a hotter system cools faster than a colder one under certain conditions~\cite{Mpemba_1969}.
In recent years, such anomalous relaxation behavior has been generalized to the quantum realm in both isolated and open quantum systems; see Ref.~\cite{Ares:2025onj} for a review.
%\textcolor{blue}{Early signatures of the quantum Mpemba effect (QME) were observed in dissipative systems undergoing first-order phase transitions~\cite{Nava_2019}, and the effect was subsequently established more generally in Markovian open quantum systems,} %The quantum Mpemba effect (QME) was initially confirmed in Markovian open quantum systems, 
% % where the mechanism was attributed to the suppressed excitation of the slowest decaying eigenmode~\cite{Carollo:2021hew}. Subsequent studies extended this concept to isolated integrable models and many-body quench dynamics~\cite{Rylands:2023yzx}.
While the classical Mpemba effect was originally discussed in terms of cooling, its quantum counterpart is usually characterized through the relaxation of a suitable distance from equilibrium~\cite{lu2017nonequilibrium}.
The quantum Mpemba effect (QME) occurs when, under the same relaxation dynamics, a state initially farther from equilibrium approaches equilibrium faster than a closer state.
In isolated many-body systems, the effect can appear as an anomalously fast restoration of a broken symmetry, e.\,g., as diagnosed by entanglement asymmetry \cite{Ares:2022koq,Rylands:2023yzx, Cao:2025oxc}.
%A recent study \cite{Cao:2025oxc} also indicates the emergence of the QME during symmetry restoration in fast scrambling systems.
In open quantum systems, the QME can be formulated in terms of relaxation under a Liouvillian, where suitable initial states have suppressed overlap with the slowest decaying mode \cite{Zhang:2024juc}.
This viewpoint is especially useful for systems whose late-time dynamics is governed by a discrete set of relaxation modes.
%\textcolor{blue}{The QME has been shown to be ubiquitous even in generic chaotic many-body systems lacking global symmetries~\cite{Bhore:2025nko}.}
%A related mechanism has also been identified in closed chaotic systems, where suppressed overlap with the slowest Ruelle–Pollicott resonance leads to anomalously fast relaxation~\cite{Yamashika:2026xpe}.
A related mechanism has also been identified in closed chaotic systems~\cite{Yamashika:2026xpe}.
%A recent study \cite{Summer:2025wsa} has also provided a resource-theoretic perspective, unifying thermal and symmetry-restoration Mpemba effects within a common framework.
%A recent study has also provided a resource-theoretic perspective on the QME~\cite{Summer:2025wsa}.
It should be noted that the characterization of the (quantum) Mpemba effect can depend on the choice of distance measures or observables used to quantify the approach to equilibrium \cite{lu2017nonequilibrium}.
This issue has motivated recent attempts to formulate the effect in a more measure-independent way, for example through thermomajorization theory \cite{VanVu:2024zps}.
%These explorations provide critical perspectives. A crucial aspect yet to be fully understood is the intrinsic connection between this anomalous relaxation and the underlying macroscopic energy dissipation mechanisms. A crucial aspect that remains to be fully understood is how this anomalous relaxation is connected to macroscopic energy dissipation.
While the QME has been studied in various quantum settings, its realization in strongly coupled nonequilibrium systems remains a natural and important direction.
%A key challenge is to understand how such anomalous relaxation emerges in strongly coupled many-body systems.
This connection becomes particularly significant when a quantum many-body system that exhibits fast scrambling and thermalization undergoes nonequilibrium relaxation accompanied by a continuous phase transition.
%\textcolor{red}{[Introduce a question in CMT side(?)]}

Holographic duality~\cite{Maldacena:1997re, Gubser:1998bc, Witten:1998qj}
%, established on the microscopic foundations of string solitons~\cite{Duff:1994an},
provides a powerful framework for studying quantum many-body systems, see review~\cite{Zaanen_Liu_Sun_Schalm_2015}.
It allows us to analyze such quantum systems by mapping them to gravity models, even if the systems are far from equilibrium~\cite{Liu:2018crr}.
%This framework is based on the geometry of the black hole horizon, which introduces natural dissipation effects to the dual boundary system~\cite{Adams:2012pj}. Black holes, acting as fast scramblers that saturate the quantum chaos bound, provide a natural geometric paradigm for characterizing the macroscopic thermalization and energy dissipation of the system.
Within the framework of holography, black holes are not only geometric objects but also thermal systems characterized by a Hawking temperature.
Furthermore, the black hole is considered as the fastest scrambler saturating the chaos bound~\cite{Maldacena:2015waa} and provides a natural geometric paradigm for characterizing thermalization and energy dissipation~\cite{Adams:2012pj,Tian:2019}.
The holographic framework has been extensively applied to study different nonequilibrium phenomena, including condensation processes~\cite{Murata:2010dx, Bhaseen:2012gg}, nonequilibrium steady states~\cite{Karch:2007pd,Nakamura:2012ae,Ali-Akbari:2013hba,Li:2013fhw,Ishii:2018ucz,Ishigaki:2020vtr}, nonequilibrium dynamical transitions~\cite{Guo:2018mip,Zeng:2018ero,Li:2020omw}, quantum turbulence~\cite{Adams:2012pj,Du:2015,Lan:2016cgl,Zeng:2024rwn}, the Kibble-Zurek mechanism~\cite{Sonner:2014tca,Chesler:2014gya,Natsuume:2017jmu,delCampo:2021rak}, time-crystal-like behaviors~\cite{Yang:2023dvk, Lei:2026znq}, pattern formation~\cite{Xia:2024ton}, and vortex dynamics~\cite{Wittmer:2020mnm,Lan:2023gyc,Lan:2023qia,Yang:2024vga,Yang:2025bsw}.
Motivated by these developments, we ask how the QME manifests itself in holographic nonequilibrium dynamics.
%The holographic framework enables an analysis of nonequilibrium dynamics in strongly coupled systems from both macroscopic and microscopic perspectives on dissipation. Investigating the QME, an anomalous dynamical phenomenon occurring during thermalization, within the holographic framework effectively characterizes this effect under quantum dissipative conditions. It leverages the geometric properties of the black hole horizon to directly reveal the internal correlation between the QME and macroscopic dissipation mechanisms from a gravitational perspective. It is of interest to understand how this anomalous relaxation is realized in the holographic gravity model.
%A recent holographic study~\cite{Benini:2024xjv} established the QME in holographic systems via entanglement asymmetry.
%A complementary holographic study on the QME was recently proposed in terms of entanglement asymmetry with perturbative analysis \cite{Benini:2024xjv}.
%To probe the microscopic origin of this effect, however, one must go beyond the static diagnosis and track the real-time relaxation.
%To obtain further understanding of this effect, however, one must go beyond the perturbative analysis and track the nonlinear real-time dynamics.
%}

\begin{figure}[h]
    \centering
    \includegraphics[width=7cm]{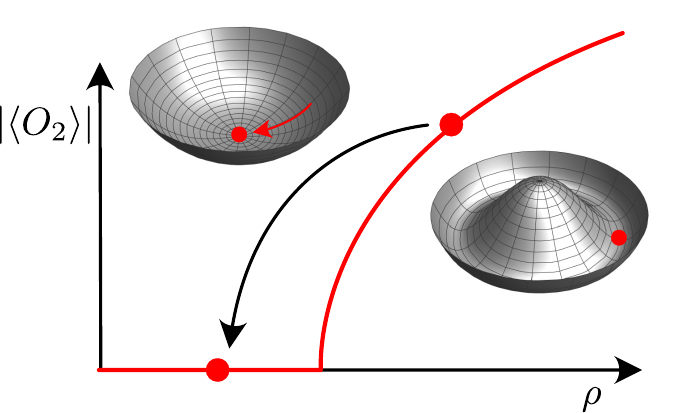}
    \caption{Schematic of our setup.
    The system is initially prepared in the symmetry-broken phase ($\rho > \rho_c$).
    After that, it relaxes to a state in the symmetry-restored phase ($\rho < \rho_c$).
    The insets sketch the corresponding effective potentials, and the curved arrows illustrate the relaxation dynamics towards the symmetry-restored equilibrium.}
    \label{fig:sketch}
\end{figure}

In this Letter, we investigate the QME in the holographic superfluid model by analyzing nonlinear real-time dynamics.
%using the order parameter and the energy dissipation into the black hole.
Starting from initial states with spontaneously broken U(1) symmetry, we compute the relaxation toward the symmetry-restored equilibrium state (see Fig.\,\ref{fig:sketch}).
We then find that the QME emerges in a wide parameter range.
%, where states with strong initial symmetry breaking relax faster toward equilibrium than those with weak breaking.
% We reveal that this anomalous relaxation behavior is intrinsically connected to energy dissipation dynamics.
Throughout the relaxation process, the order parameter may temporarily increase while the shifted free energy decreases monotonically, which is computed from the energy flux into the black hole.
This behavior implies that the shifted free energy serves as a natural distance measure from equilibrium.
%Furthermore, we attempt to understand the QME in terms of quasinormal modes (QNMs) of the final state.
We further develop a QNM-based interpretation of the QME.
Extracting the contribution of each QNM throughout the relaxation process, we find a suppression of the slowest mode and amplification of the second mode within the parameter region where the QME occurs.
These results imply that the QME is governed by the dynamical competition among the QNMs in this holographic superfluid model.
Our analysis provides a holographic understanding of the QME in nonequilibrium dynamics involving strongly coupled degrees of freedom.

\prlsection{Holographic setup}
We employ the standard holographic superfluid model \cite{Hartnoll:2008vx, Hartnoll:2008kx}, whose action is given by
\begin{equation}
    S = \int\dd[4]{x}\sqrt{-g}\left[
        - \frac{1}{4} F^2
        - \abs{D_{\mu} \Phi}^2
        - m^2 \abs{\Phi}^2
    \right],
\end{equation}
where $F_{\mu\nu} = \partial_{\mu} A_{\nu} - \partial_{\nu} A_{\mu}$ is the field strength of the $U(1)$ gauge field $A_{\mu}$, $\Phi$ is a complex scalar field, and $m$ is the bulk mass.
%, which we will set $m^2 = -2$.
$D_{\mu} = \nabla_{\mu} - i q A_{\mu}$ is the gauge-covariant derivative.
In this study, we focus on the probe limit assuming that the backreaction of the matter fields on the background spacetime is negligible.
We employ the Schwarzschild-AdS$_4$ black brane as the background.
In the ingoing Eddington-Finkelstein coordinates, the metric is given by
\begin{equation}
    \dd{s}^2 = \frac{- f(z) \dd{t}^2 - 2 \dd{t}\dd{z} + \dd{x}^2 + \dd{y}^2}{z^2},
\end{equation}
where $f(z) = 1 - \frac{z^3}{z_{h}^3}$.
The black hole horizon is located at $z=z_{h}$, while the AdS boundary is located at $z=0$.
The Hawking temperature is given by $T = \frac{3}{4\pi z_{h}}$.
The equations of motion are given by
\begin{align}
	\nabla_{\mu} F^{\mu\nu}
	- i q (\Phi^{*} D^{\nu} \Phi - \Phi D^{\nu*} \Phi^{*}) =& 0,\\
	D_{\mu} D^{\mu} \Phi - m^2 \Phi =& 0.
\end{align}
The bulk energy-momentum tensor associated with the Maxwell-scalar theory is given by
\begin{equation}
\begin{aligned}
    T_{\mu\nu} =&
    (D_{\mu}\Phi)^{*} (D_{\nu}\Phi)
    + (D_{\nu}\Phi)^{*} (D_{\mu}\Phi)\\
    -& g_{\mu\nu} \left(
        |D \Phi|^2 + m^2 |\Phi|^2    
    \right)
    + F_{\mu\rho} F_{\nu}{}^{\rho}
    - \frac{1}{4} g_{\mu\nu} F^2.
\end{aligned}
\label{eq:energy-momentum_tensor}
\end{equation}

%In this study, we do not consider $x$- and $y$-dependence so the fields are functions of $t$ and $z$. 

For simplicity, we focus on a spatially homogeneous and isotropic configuration, meaning that all bulk fields depend only on the time $t$ and the holographic radial coordinate $z$. We do not consider the $x$- and $y$-components of the vector fields, either.
By choosing the radial gauge $A_{z}=0$, only $A_{t}$ is relevant in our setup.
In this study, we set $m^2 = -2$ and $q=1$.
For the choice of the bulk mass, the scaling dimensions of the scalar operator are determined as $\Delta_{-} = 1$, and $\Delta_{+} = 2$.
We can write the asymptotic expansion of the scalar field as
\begin{equation}
    \Phi = z \Phi_{(1)} + z^2 \Phi_{(2)} + \order{z^3},
\end{equation}
and for the $t$-component of the vector field
\begin{equation}
    A_{t} = \mu - \rho z + \order{z^2},
\end{equation}
where $\mu$ and $\rho$ are the chemical potential and the charge density, respectively.
For the scalar field, we impose the Dirichlet boundary condition $\Phi_{(1)} = 0$ and identify $\Phi_{(2)}$ as the scalar condensate $\expval{O_{2}}$ with the scaling dimension $\Delta = 2$, which is an order parameter associated with the $U(1)$ symmetry.
For the vector field, we impose the Neumann boundary condition at $z=0$, which corresponds to the fixed charge density.

In this study, we set $z_{h}=1$ corresponding to $T=\frac{3}{4\pi}$, for simplicity.
By using the scaling symmetry, all quantities can be interpreted as being measured in units of appropriate powers of $T$.
Under the vanishing Dirichlet condition on the scalar field, the system exhibits spontaneous $U(1)$ symmetry breaking for large $\rho>\rho_{\rm c}$, where the critical density is $\rho_{\rm c} = 4.06$.

\prlsection{Time evolution of the order parameter}
To investigate the QME in our holographic model, we compute the time evolution of the order parameter under a specific value of the charge density $\rho = \rho_{\rm f} < \rho_{\rm c}$ where the $U(1)$ symmetry is restored.
%The initial states at $t=0$ are prepared as stationary solutions under another value of $\rho = \rho_{\rm i} > \rho_{\rm c}$ where the $U(1)$ symmetry is spontaneously broken.
We prepare the initial states at $t=0$ as stationary solutions under another value of $\rho = \rho_{\rm i} > \rho_{\rm c}$ where the $U(1)$ symmetry is broken spontaneously.
The setup is equivalent to considering an instantaneous quench from $\rho = \rho_{\rm i}$ to $\rho_{\rm f}$ at $t=0$.
%For further details of the numerical calculation, see Appendix \ref{appendix:details}.
Further details of the numerical method are given in the Appendix.
% \begin{equation}
%     \rho(t) =
%     \begin{cases}
%         \rho_{\rm i} & t \leq 0\\
%         \rho_{\rm f} & 0 < t
%     \end{cases}.
%     \label{eq:density_quench}
% \end{equation}
% {\color{magenta}%
% However, the time dependence of the charge density must obey the conservation law:
% \begin{equation}
%     - \partial_{t} \rho(t) = i \left(
%         \phi_{\rm s}^{*} \expval{O_{2}} - \phi_{\rm s} \expval{O_{2}^{*}}
%     \right),
% \end{equation}
% where $\phi_{\rm s}$ is an external source to the scalar operator.
% The conservation law is obtained as the gauge constraint in the bulk.
% Substituting Eq.~(\ref{eq:density_quench}) into the conservation law, we deduce
% \begin{equation}
%     \phi_{\rm s}(t) \propto i \frac{\rho_{\rm f} - \rho_{\rm i}}{2q \abs{\expval{O_{2}(0)}}} \times \delta(t).
% \end{equation}
% We assumed $\expval{O_{2}(t)}$ is smooth at $t=0$ and set $\Im \expval{O_{2}(0)} = 0$ by virtue of the $U(1)$ symmetry.
% Therefore, the instantaneous quench of the density (\ref{eq:density_quench}) effectively induces the delta-peak perturbation of the external source at $t=0$.
% }%

\begin{figure}[htp!]
\centering
\includegraphics[width=8cm]{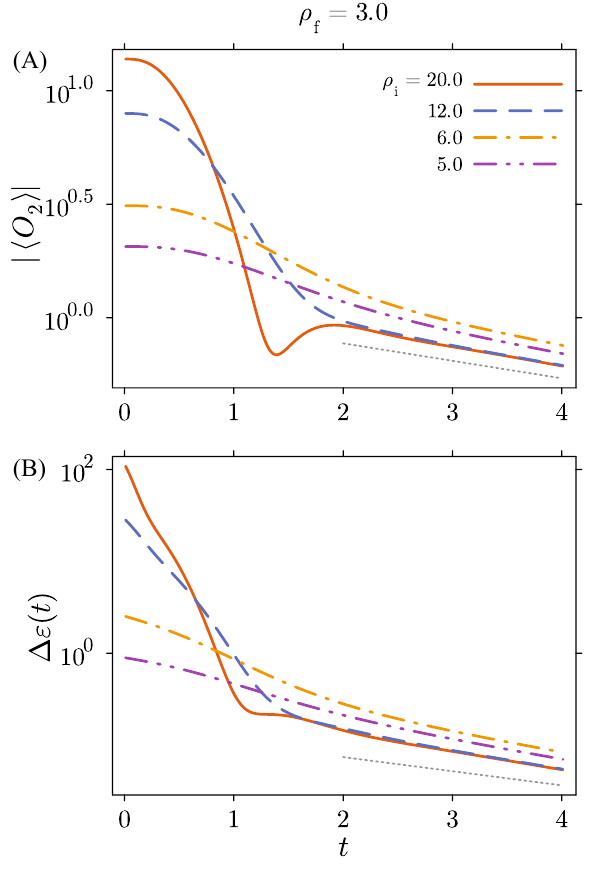}
\caption{%
    (A) Time evolution of the condensate for various $\rho_{\rm i} \in \{5.0, 6.0, 12.0, 20.0\}$.
    During the relaxation process, the charge density is fixed at $\rho_{\rm f} = 3.0$.
    %Note that the vertical axis is shown in log scale.
    The gray dotted line shows the behavior of the slowest decaying mode. 
    (B) Shifted free energy as a function of time.
}
\label{fig:O2_and_dissipation}
\end{figure}

Figure \ref{fig:O2_and_dissipation}(A) shows the time evolution of the condensate for various values of the initial charge density from $\rho_{\rm i} = 5.0$ to $20.0$.
Here, we set the (post-quench) charge density $\rho_{\rm f} = 3.0$ for $t\geq 0$.
By comparing the results for $\rho_{\rm i} = 5.0$ and $6.0$, it appears at first glance that the larger the initial condensate, the longer it takes to reach equilibrium.
However, for instance, the result for $\rho_{\rm i} = 12.0$ shows more rapid decay than the other cases, even though its initial condensate is large.
This behavior explicitly indicates the emergence of the QME in the holographic superfluid model.
At sufficiently late times, the behavior of the condensate is governed by the slowest QNM in the final state, regardless of the initial states~\cite{Murata:2010dx, Bhaseen:2012gg}.
In Fig.~\ref{fig:O2_and_dissipation}(A), we show the behavior of $e^{-\Im \omega_{1} t}$ as the gray dotted line, where $\omega_{1} = 0.3726 - 0.1771i$ is the slowest QNM frequency in the final state.
From $t\approx 2$, the curves exponentially decay with this decay rate.

%As a distance measure from equilibrium, however, the condensate, which is an order parameter of the $U(1)$ symmetry, is not a suitable choice.
%Meanwhile, the condensate, which is an order parameter of the $U(1)$ symmetry, is not a suitable choice as a distance measure from equilibrium.
%From Fig.~\ref{fig:O2_and_dissipation}(A), the condensate temporally increases around $t\approx 1.5$ for $\rho_{\rm i} = 20.0$, so it is nonmonotonic.
%A distance measure should decrease monotonically throughout the relaxation process~\cite{lu2017nonequilibrium}.
%In our case, the energy dissipation satisfies this requirement, as follows.
%In the holographic superfluid model under the probe limit, the generalized free energy studied in \cite{Tian:2019} satisfies this requirement.

%We now examine the time dependence of the generalized free energy, where the QME occurs.
Next, we verify the QME using the shifted free energy introduced below.
The energy dissipation rate in the boundary system can be holographically evaluated from the energy-momentum tensor (\ref{eq:energy-momentum_tensor}) at the horizon~\cite{Adams:2012pj,Tian:2019}.
In our case, we obtain
\begin{equation}
\begin{aligned}
    \dv{\varepsilon}{t} =& \left.\sqrt{-g} T^{z}{}_{t}\right|_{z = z_{h}}
    = - 2 z_{h}^{-2} \abs{\partial_{t} \Phi (z_{h})}^2,
\end{aligned}
\label{eq:energy_flux}
\end{equation}
where $\varepsilon$ is the generalized free energy density \cite{Tian:2019} of the system.
The energy dissipation rate is the negative of the above quantity.
Integrating over time, we obtain
\begin{equation}
    \Delta \varepsilon(t) =
    \lim_{t_{\rm f} \to \infty}
    \int_{t_{\rm f}}^{t} \dd{t'}
    \left.\sqrt{-g} T^{z}{}_{t}\right|_{z=z_{h}},
    \label{eq:energy_dissipation}
\end{equation}
which is actually the generalized free energy shifted so that $\Delta \varepsilon(\infty)=0$.
We refer to $\Delta\varepsilon(t)$ as a shifted free energy.
Since Eq.~(\ref{eq:energy_flux}) is always negative, $\Delta \varepsilon(t)$ is a monotonically decreasing function of $t$ throughout the relaxation process.
Such a quantity is more suitable as a distance measure from equilibrium \cite{lu2017nonequilibrium}.

Figure \ref{fig:O2_and_dissipation}(B) shows the shifted free energy as a function of time for various values of the initial density.
Here we set $t_{\rm f} = 10.0$ to perform the time integration in Eq.~(\ref{eq:energy_dissipation}) numerically.
Since we impose the Dirichlet boundary condition $\Phi_{(1)}(t) =0$ and the fixed charge density $\rho(t) = \rho_{\rm f}$ for $t>0$, we can regard all the energy as being injected at $t=0$.
Similar to the behavior of the condensate, the shifted free energy quickly decays at intermediate times for $\rho_{\rm i} = 12.0$ and $20.0$, where the QME occurs.
For $\rho_{\rm i} = 20.0$, the shifted free energy shows a short plateau around $t\approx 1.2$ but it is actually monotonically decreasing.
From this figure, it is clear that the shifted free energy for $\rho_{\rm i} = 20.0$ has two distinct time domains before and after the plateau.
After the plateau, the shifted free energy is also governed by the slowest mode.
In Fig.~\ref{fig:O2_and_dissipation}(B), we show the behavior of $e^{-2\Im\omega_{1} t}$ as the gray dotted line.
%, where $\omega_{1}$ is the frequency of the slowest QNM given by the inset table of Fig.~\ref{fig:QNM_profiles}. %% repeated
Note that the decay rate is expected as $-2\Im\omega_{1}$ from Eq.~(\ref{eq:energy_flux}).

\begin{figure}[htbp]
\centering
\includegraphics[width=8cm]{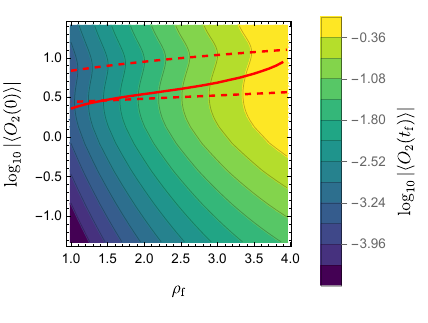}
\caption{%
    Late-time condensate $\abs{\expval{O_{2}(t_{\rm f})}}$, at $t_{\rm f} = 10.0$, as a function of the initial condensate $\abs{\expval{O_{2}(0)}}$ and $\rho_{\rm f}$.
    %Note that the plotted values of the condensates are $\log_{10}\abs{\expval{O_{2}(t_{\rm f})}}$ and $\log_{10}\abs{\expval{O_{2}(0)}}$.
    %The initial condensate $\abs{\expval{O_{2}(0)}}$ is parameterized by $\rho_{\rm i}$.
    The red dashed curves denote $\pdv{\abs{\expval{O_{2}(t_{\rm f})}}}{\abs{\expval{O_{2}(0)}}} = 0$.
    %The QME is mainly observed in the region between these two curves.
    The red solid curve denotes crossing points between $w_{1}$ and $w_{2}$ defined by Eq.~(\ref{eq:time_integrated_QNM_amplitude}).
}
\label{fig:O2_contour}
\end{figure}

We now evaluate the amount of the QME
%quantitatively,
by using the late-time condensate $\abs{\expval{O_{2}(t_{\rm f})}}$.
At sufficiently late times $t_{\rm f}$, the time evolution of the condensate is governed by the slowest decaying mode, so the relative amount of the late-time condensate is mainly determined during the early-time relaxation.
%We also discard the early-time behavior to avoid detecting the oscillating behavior for large initial values.
% Figure \ref{fig:O2_late_time} shows the late-time condensate $\abs{\expval{O_{2}(t_{\rm f})}}$ as a function of the amplitude of the initial condensate $\abs{\expval{O_{2}(0)}}$ for $\rho_{\rm f} = 2.5$, $3.0$, and $3.5$.
Figure \ref{fig:O2_contour} shows the late-time condensate as a function of the charge density and the initial condensate.
Here, we set the time for measuring the late-time condensate to $t_{\rm f} = 10.0$.
The initial condensate is parameterized by the initial charge density $\rho_{\rm i}$.
Due to the critical slowing down, the late-time condensate generally increases as $\rho_{\rm f}$ approaches $\rho_{\rm c} = 4.06$.
For sufficiently small initial condensates, the late-time condensate increases monotonically along the vertical axis, but turns over and begins to decrease beyond a certain point.
In Fig.~\ref{fig:O2_contour}, we show the locations of such points where $\pdv{\abs{\expval{O_{2}(t_{\rm f})}}}{\abs{\expval{O_{2}(0)}}} = 0$ by the red-dashed curves.
For $\rho_{\rm f} = 3.0$, we can observe the local peak of the late-time condensate at $\abs{\expval{O_{2}(0)}} \approx 3.4 = 10^{0.53}$.
This behavior corresponds to the emergence of the QME.
For $\rho_{\rm f} = 3.0$, the QME starts to become pronounced around $\abs{\expval{O_{2}(0)}} \approx 3.4$ and disappears around $\abs{\expval{O_{2}(0)}} \approx 10$.

%From this figure, one can see the emergence of the QME in this model more comprehensively.

\prlsection{Model fit with quasinormal modes}
%The late-time behavior of the condensate is described by the exponential decay, whose decay time is given by the slowest QNM in the finial equilibrium state, regardless of the initial states~\cite{Murata:2010dx, Bhaseen:2012gg}.
Although our system is nonlinear, we can still understand the behavior of the condensate and the QME by using the QNMs approximately.
The QNMs can be analyzed by considering a small fluctuation around a background solution.
We can expect that the QNMs, which are associated with the perturbation of the scalar field $\Phi$ around the final state, will be related to the relaxation process.
The QNMs in this system have been studied in \cite{Amado:2009ts}.
Here, we adopt a natural and convenient scheme first applied in \cite{Du:2016} to numerically compute QNMs, as described in Appendix.

% \begin{table}[htp!]
% \centering
% \caption{%
% Frequencies of the first four QNMs associated with the perturbation of $\Phi$ around the symmetry restored phase for $\rho_{\rm f} = 3.0$.
% The mode frequencies associated with $\Phi^{*}$ are given by $-\omega_{n}^{*}$.
% }
% \label{table:QNMs}
% \begin{tabular}{@{}|l|c|c|@{}}
% \hline
% $n$ & $\Re \omega_{n}$ & $\Im \omega_{n}$ \\\hline\hline
% $1$ & $0.372618$ & $- 0.177123 $\\
% $2$ & $1.58657 $ & $- 1.76571 $\\
% $3$ & $2.76564 $ & $- 3.78189 $\\
% $4$ & $-3.45548$ & $- 3.77226$ \\\hline
% \end{tabular}
% \end{table}
\begin{figure}
    \centering
    \includegraphics[width=8cm]{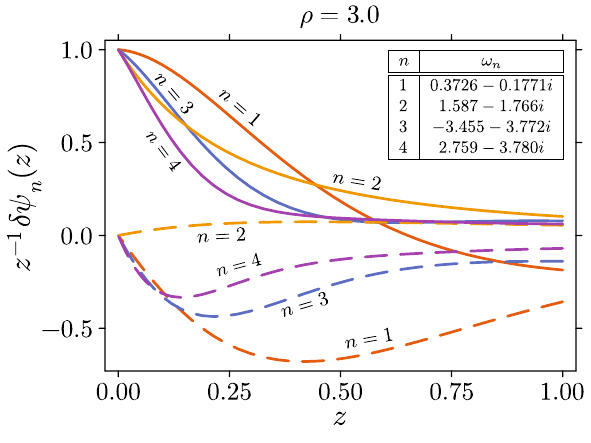}
    \caption{%
        Bulk profiles of the first four QNMs for $\rho=3.0$.
        The solid curves denote the real part while the dashed curves the imaginary part. 
        The inset table shows corresponding QNM frequencies.
    }
    \label{fig:QNM_profiles}
\end{figure}

\begin{figure}
\centering
\includegraphics[width=8cm]{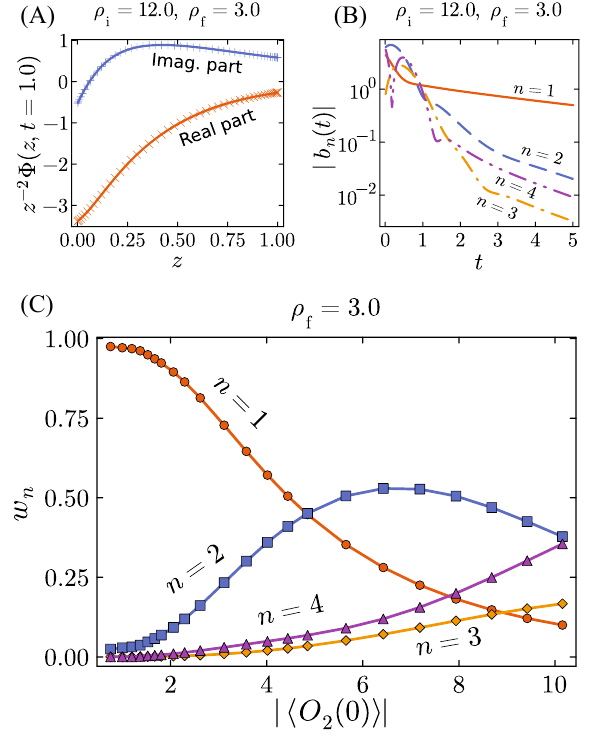}
\caption{%
    (A) Fitted curves of the bulk scalar field for $\rho_{\rm i} = 12.0$ and $\rho_{\rm f} = 3.0$, at $t=1.0$.
    The points denote the data, while the (mostly overlapping) curves denote the fit result.
    (B) Time evolution of the QNM amplitudes.
    (C)
    QNM weight as a function of the initial condensate for $\rho_{\rm f} = 3.0$.
}
\label{fig:QNM_fit}
\end{figure}

When the system is close to equilibrium, a small perturbation field can be written as a linear combination of the QNMs.
%Extending the analysis in a linear regime to a nonlinear regime, we fit the bulk profile of the scalar field using the following model:
Motivated by the linear QNM expansion near equilibrium, we fit the nonlinear bulk profile of the scalar field using the following model:
\begin{equation}
    \Phi(t, z) \simeq z \sum_{n=1}^{N_{\rm max}} b_{n}(t) \var{\psi}_{n}(z),
    \label{eq:radial_QNM_expansion}
\end{equation}
where $\var{\psi}_{n}(z)$ is the $n$-th QNM function, and $b_{n}(t)$ is a complex-valued fitting parameter for each $t$.
The QNM amplitude is represented by $\abs{b_{n}}$.
%Here, we fix the normalization of the QNMs by $\lim_{z\to 0} \partial_{z} \var{\psi}_{n}(z) = 1$, then the value of $b_{n}(t)$ is also fixed.
We fix the QNM normalization by imposing $\lim_{z\to 0} \partial_{z} \var{\psi}_{n}(z) = 1$, which uniquely determines the corresponding coefficients $b_{n}(t)$.
Since we are interested in the mode competition between the lowest-lying QNMs, we set $N_{\rm max} = 4$ and use the first four QNMs shown in Fig.~\ref{fig:QNM_profiles} for $\rho = 3.0$.%
\footnote{
    For other values of $\rho = \rho_{\rm f}$, we keep the label $n$ by tracking $\omega_{n}$.
}
%, whose frequencies and residues are listed in Table \ref{table:QNMs} for $\rho_{\rm f} = 3.0$.%
The higher-order QNMs with large imaginary parts of the frequency decay rapidly in a short time, so we can omit the contributions of the higher QNMs.
%In addition, the higher QNMs usually have a large real-part of the frequency implying that such modes requires high energy to excite.
%The calculation of the higher-order QNMs also requires higher numerical precision.
%For the reasons stated above, we can omit the contributions of the higher QNMs.
%We expect that the lowest few QNMs dominate the intermediate-to-late dynamics.
Note that the expansion (\ref{eq:radial_QNM_expansion}) is not generally justified even for $N_{\rm max}=\infty$ since QNMs do not form a complete basis.
Figure \ref{fig:QNM_fit}(A) shows the fitted curves of $\Phi(t,z)$ for $\rho_{\rm f} =3.0$ and $\rho_{\rm i} = 12.0$ at an intermediate time $t=1.0$.
%Despite the data being a nonlinear solution, the QNM expansion approximates it very well.
Surprisingly, the model fits the nonlinear result of $\Phi(t,z)$ almost perfectly, throughout the relaxation process.
In other words, the nonlinear effect is successfully encoded into $b_{n}(t)$ and interactions among the QNMs.
Figure \ref{fig:QNM_fit}(B) shows the corresponding time evolution of $\abs{b_{n}(t)}$.
As expected, the coefficient $b_{n}(t)$ no longer behaves as $e^{-i\omega_{n} t}$ in the linear case.
Furthermore, the higher-mode coefficients are amplified at early times, reflecting the nonlinear dynamics.
This behavior becomes pronounced when the initial condensate is sufficiently large and the QME occurs, such as for $\rho_{\rm i} = 12.0$.

To evaluate the contribution of the QNMs throughout the process and compare the results across different initial values, we introduce the following weight
\begin{equation}
	w_{n} := \frac{
		\int_{0}^{t_{\rm f}}\dd{t} \abs{b_{n}(t)}^2
	}{
		\sum_{n=1}^{N_{\rm max}} \int_{0}^{t_{\rm f}}\dd{t} \abs{b_{n}(t)}^2
	}.
    \label{eq:time_integrated_QNM_amplitude}
\end{equation}
The $n$-independent part is canceled out by this definition.
Using this quantity, we can estimate the contribution of each QNM throughout the relaxation process.
Figure \ref{fig:QNM_fit}(C) shows the QNM weight (\ref{eq:time_integrated_QNM_amplitude}) as a function of the initial condensate.
From this result, the slowest ($n=1$) mode is dominant when the initial condensate is sufficiently small.
However, as the initial condensate increases, the contribution of the second ($n=2$) mode increases.
%The second mode is dominant at $\abs{\expval{O_{2}(0)}} \approx 3.58$, which roughly agrees with the peak of the late-time condensate as shown in Fig.~\ref{fig:O2_late_time}, and the slowest first mode is suppressed.
This observation provides an understanding of the mechanism of the QME from the aspect of the holography: The contribution of the slowest QNM is suppressed while the second QNM is enhanced, resulting in the QME.
For $\rho_{\rm f} = 3.0$, the contributions of the slowest mode and the second mode is comparable around $\abs{\expval{O_{2}(0)}}\approx 5.0$, i.e., $w_{1}=w_{2}$.
In Fig.~\ref{fig:O2_contour}, we show the locations of such crossing points by the solid red curve for various $\rho_{\rm f}$.
The crossing points are located within the QME region, where $\pdv{\abs{\expval{O_{2}(t_{\rm f})}}}{\abs{\expval{O_{2}(0)}}} < 0$, for almost all $\rho_{\rm f}$.
This observation supports our understanding of the QME mechanism in this system.
%As expected, the slowest mode is dominant for $\rho_{\rm i} = 5.0$ where the initial condensate is relatively small.
%For large initial condensates, the amplitudes of the higher modes are enhanced.
%This means the initial state have large overlap with the higher modes.
%Furthermore, in the initial stages of relaxation, the amplitude of higher modes increases.
%We may understand that such behavior is caused by the nonlinear dynamics.

\prlsection{Conclusion}
In this work, we have investigated the relaxation dynamics of the holographic superfluid to examine the emergence of the QME during the $U(1)$ symmetry restoration process and to understand its realization mechanism in holography.
The results confirm that initial states with stronger symmetry breaking can relax to thermal equilibrium more rapidly, which implies the QME.
By analyzing the energy flux through the black hole horizon in the bulk, we have also revealed the connection between the QME and the shifted free energy.
The shifted free energy satisfies the requirements of a distance measure: a distance measure should be a monotonically decreasing function throughout the process. 
%From a macroscopic viewpoint, QME is governed by the shifted free energy through the black hole horizon. Different initial symmetry breaking dictate varying rates of energy flux into the horizon and accelerates the overall thermalization.
On the other hand, the QME can be understood in terms of the QNMs.
We have extracted the QNM contribution at each time by considering a fitting model that employs radial QNM functions.
With this approach, the nonlinear dynamics is encoded in the time-dependent amplitudes of QNMs, reflecting interactions among the modes.
%From the fit result, we extract the contribution of each QNM throughout the relaxation process.
Comparing the contributions of the QNMs, we find that suppression of the slowest mode and amplification of the second mode occur in the parameter region where the QME occurs.
The mechanism revealed in our study aligns with the common understanding in open quantum systems that the QME emerges from the selective suppression of the normally dominant slowest decaying mode, forcing the system to relax through faster decaying channels.
A crucial difference between our holographic model and the Lindblad systems is that our model is nonlinear and the amplitudes of the QNMs are determined as a result of the nonlinear dynamics.
%\sout{Our analysis also shows that early nonlinear evolution plays an important role in setting the initial amplitudes of these modes.}
%By combining macroscopic dissipation with microscopic mode competition, our findings offer a clear physical picture for the origin of anomalous relaxation in quantum many body systems far from equilibrium.
Our findings offer a clear physical picture for the origin of anomalous relaxation in quantum many body systems far from equilibrium.

%{\color{magenta}
As demonstrated, the late-time dynamical evolution is consistently governed by the slowest decaying QNM, whose lifetime is predominantly dictated by the black hole temperature. Specifically, for $\rho_{\rm f} = 3.0$, the maximum dissipation time evaluates to $\tau_{1} = 1/\abs{\Im\omega_{1}} = 5.64$. While the temperature was held constant throughout our numerical analysis, scale invariance implies that the physical relaxation time scales as $\tau_{1} \sim 1.34/T$. Broadly speaking, this characteristic timescale aligns with the bounds of Planckian dissipation~\cite{zaanen2004temperature, Hartnoll:2021ydi}, $\tau_{\rm Pl} \sim 1/T$ in our convention $\hbar = k_{\rm B} = 1$.
%}

Future research can expand on this study in several directions.
%It is of great interest to examine the universality of the QME in holographic systems with other symmetry breaking, such as $Z_2$ symmetry.
It would be of interest to examine whether the QME occurs in holographic systems with other symmetry-breaking patterns.
%, such as $Z_2$ symmetry.
%For instance, the $Z_{2}$ symmetry breaking can be realized by considering double-trace deformation in holography~\cite{Faulkner:2010gj}.
It is also interesting to examine the QME in holographic systems showing a first-order phase transition, such as \cite{Franco:2009yz, Franco:2009if}.
From a study using the Lindblad equation \cite{Nava_2019}, we can expect that the phase coexistence will play an important role in the QME in such a setup.
%The present study has been carried out in the probe limit, while future work that includes full backreaction from the gravitational sector will refine the physical picture.
%Connecting these holographic predictions with experimental observations of quench dynamics in quantum platforms, especially superfluids and superconductors, is a key step toward laboratory verification.
%Integrating the study of the QME with broader nonequilibrium nonlinear phenomena, such as pattern formation and topological defect dynamics, will provide new insights into the complex dynamics of quantum systems far from equilibrium.
An important direction for future work is to connect these holographic results with experimental observations of quench dynamics in quantum platforms, particularly superfluids and superconductors.
%Another promising avenue is to investigate the QME in relation to broader nonequilibrium nonlinear phenomena, such as pattern formation and topological defect dynamics.
%These connections may offer new insights into the complex dynamics of quantum systems far from equilibrium.

%\begin{acknowledgments}
\emph{Acknowledgments}---
We would like to thank Hisao Hayakawa, Hua-Bi Zeng, Peng Yang and Jia Du for helpful discussions.
SI thanks the Yukawa Institute for Theoretical Physics at Kyoto University for its hospitality during the workshop "Quantum Thermalization, Hydrodynamics and Gravity."
XHG was partially supported by the National Natural Science Foundation of China (NSFC) (Grant Nos.\,12275166 and 12311540141). YT was partially supported by NSFC, China (Grant Nos.\,12375058 and 12361141825). SI was partially supported by NSFC, China (Grant No.\,W2433015). YQL was partially supported by NSFC, China (Grant No.~12405072) and China Postdoctoral Science Foundation (Grant No.\,2024M761914).
%\end{acknowledgments}

%% From revtex4-2, the .bst file is automatically selected!!
%\bibliographystyle{apsrev4-2}
\bibliography{main}

\appendix

\section{Details of the analysis}\label{appendix:details}
\subsection{Equations of motion}
Rewriting $\Phi(t,z) = z \psi(t,z)$, we obtain the scalar equation of motion as%
\footnote{
    We use the notation $\square_{,\mu} := \partial_{\mu} \square$.
}
\begin{equation}
\begin{aligned}
\mathcal{F}_{\psi} :=&
- \psi_{,zt}
+ i q A_{t} \psi_{,z}
+ \frac{1}{2}f(z) \psi_{,zz}\\
+& \frac{1}{2}f'(z) \psi_{,z}
- \frac{1}{2} z \psi
+ \frac{1}{2} i q\psi A_{t,z} = 0.
\end{aligned}
\end{equation}
For the scalar field, we impose the boundary condition by $\psi(t, 0)=0$ corresponding to $\Phi_{(1)}(t) = 0$.
In the radial gauge $A_{z}=0$ in the ingoing Eddington-Finkelstein coordinate, the $t$ and $z$ components of the Maxwell equation become
\begin{align}
    \mathcal{C}_{1} :=&
    -A_{t,zz}
    %+ \partial_{z}(\partial_{x} A_{x} + \partial_{y} A_{y})
    + i q(\psi^{*}\partial_{z} \psi - \psi \partial_{z} \psi^{*}) = 0,\\
    \mathcal{C}_{2} :=&
    -\frac{1}{2} A_{t,zt}
	- \frac{iq}{2} (\psi^{*}\partial_{t} \psi - \psi \partial_{t} \psi^{*})\nonumber\\
	+& \frac{iq}{2} f(z) (\psi^{*}\partial_{z} \psi - \psi \partial_{z} \psi^{*})
	- q^2 A_{t} \psi^{*}\psi = 0,
\end{align}
respectively.
Note that there is a residual gauge symmetry
\begin{equation}
    A_{t} \to A_{t} + \partial_{t} \Lambda(t),\quad
    \psi \to \psi e^{iq \Lambda(t)},
    \label{eq:residual_gauge}
\end{equation}
where $\Lambda(t)$ is an arbitrary function of $t$.
%We have set $q=1$.
From $\mathcal{C}_{1} = 0$, we obtain $A_{t}$ as
\begin{equation}
    A_{t}(t, z) = \int_{z}^{z_{h}}\dd{z'}\left(
        \int_{0}^{z'}\dd{z''}
        i q (\psi^{*} \partial_{z} \psi - \psi \partial_{z} \psi^{*})
        - \rho
    \right).
\end{equation}
%Note that we imposed $A_{t}(z=z_{h}) = 0$ to fix the residual gauge (\ref{eq:residual_gauge}).
Note that we have fixed the integration constant by imposing $A_{t}(z=z_{h}) = 0$, which fixes the residual gauge (\ref{eq:residual_gauge}).
With this boundary condition, the phase of the order parameter becomes a time-independent constant in stationary solutions.
Another constraint, $\mathcal{C}_{2}=0$, is not fully independent of the other equations of motion due to the $U(1)$ gauge symmetry of the system, which actually leads to two possible evolution schemes (see, e.g. Appendix of \cite{Du:2015}). Explicitly, these equations of motion are related by \cite{Zeng:2018ero}
\begin{equation}
	\dv{z} \mathcal{C}_{2} = \dv{t} \mathcal{C}_{1}
	%- 2 \dv{x} \mathcal{F}_{x}
	%- 2 \dv{y} \mathcal{F}_{y}
	- 2 \frac{i q}{z} (\psi^{*} \mathcal{F}_{\psi} - \psi \mathcal{F}_{\psi}^{*}).
\end{equation}
Thus, as one choice of evolution scheme, it is sufficient to impose $\mathcal{C}_{2}=0$ at $z=0$.
We obtain
%\begin{equation}
%    \left.\mathcal{C}_{2}\right|_{z=0}
%    = - \frac{1}{2} A_{t,zt} = \frac{1}{2} %\partial_{t} J_{t} = 0.
%\end{equation}
\begin{equation}
\begin{aligned}
    2 \left.\mathcal{C}_{2}\right|_{z=0}
    =& \lim_{z\to 0}
        - A_{t,zt}\\
        +& i q \left(
            \psi^{*} (\partial_{z} - \partial_{v}) \psi
            -
            \psi (\partial_{z} - \partial_{v}) \psi^{*}
        \right)\\
        -& 2 q^2 A_{v} \psi^{*}\psi
     = 0.
    \label{eq:constraint_2}
\end{aligned}
\end{equation}
In our coordinate and gauge fixing, we can read the scalar source and the condensate by
\begin{align}
    \psi_{\rm s} = \lim_{z \to 0} \psi,\quad
    \expval{O_{2}} = \lim_{z \to 0}
    \psi_{,z} - \psi_{,t} + i q A_{t} \psi,
\end{align}
respectively.
Note that the term $\psi_{,t}$ comes from the difference between the iEF coordinate and the usual Poincar\'e coordinate, and the term including $A_{t}$ comes from the gauge transformation to recover the radial gauge in the Poincar\'e coordinate.
%after the coordinate transformation.
Using these relations, we can write Eq.~(\ref{eq:constraint_2}) as
\begin{equation}
    \partial_{t} \rho = i q (
        \psi_{\rm s}^{*}\expval{O_{2}}
        - \psi_{\rm s}\expval{O_{2}^{*}}
    ).
\end{equation}
Thus, the constraint (\ref{eq:constraint_2}) corresponds to the generalized Ward identity in the boundary theory.
Since we impose $\psi_{\rm s} = 0$, this equation becomes the charge conservation law and is manifestly satisfied in our setup with the fixed charge density.

In our numerical analysis, we employ the Chebyshev pseudospectral method for the differentiation and integration in the $z$ direction.
The number of grid points is $N_{z} = 50$.
For the time evolution, we use the fourth-order Runge-Kutta method with a time step $\Delta t = 0.01$.
The numerical integration for $\Delta \varepsilon(t)$ is performed using the trapezoidal rule. 

% We show the further results of the time evolution of the order parameter and the energy dissipation rate in Fig.~\ref{fig:O2_evolution_various} and Fig.~\ref{fig:dissipation_rate_various}.
% Fig.~\ref{fig:O2_evolution_large_various} and \ref{fig:dissipation_rate_large_various} show the further results for large $\rho_{\rm i}$.
% From Fig.~\ref{fig:dissipation_rate_large_various} for large $\rho_{\rm i}$, we can observe that the energy-decay rate is dominated by the decay rate of the QNMs in each time regime.

% \begin{figure*}
% \centering
% \subfloat[]{%
%     \includegraphics[width=8cm]{./fig/O2_evolution_large_various.pdf}
%     \label{fig:O2_evolution_large_various}
% }
% \quad
% \subfloat[]{%
%     \includegraphics[width=8cm]{./fig/energy_difference_large_various.pdf}
%     \label{fig:energy_difference_large_various}
% }
% \caption{%
%     (a) Time evolution of the order parameter for various
%     $\rho_{\rm i} = 5.0$ to $20.0$.
%     The dotted line denotes the slowest decaying mode in the final state.
%     (b) Dissipated energy as a function of time for various $\rho_{\rm i}$.
% }
% \label{fig:various_O2_and_dissipation}
% \end{figure*}

\subsection{Stationary solutions}

\begin{figure}
\centering
\includegraphics[width=8cm]{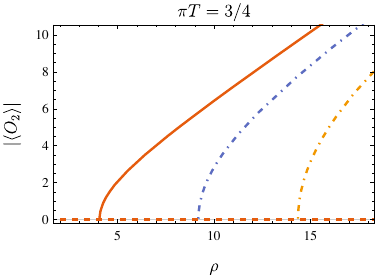}
\caption{%
	Condensate as a function of the charge density for the stationary solutions.
    The solid curve corresponds to the broken phase, whereas the dashed line on the horizontal axis corresponds to the restored phase.
    The dot-dashed curves correspond to the (unstable) higher excited states.
}
\label{fig:O2-rho}
\end{figure}

Assuming that all the fields are independent of $t$, we obtain the equations of motion for the stationary solutions.
Writing $\psi(z) = \sigma(z) e^{i\theta(z)}$, we obtain
\begin{align}
	&\sigma''(z) + \frac{f'(z)}{f(z)} \sigma'(z)\nonumber\\
    &+ \left(
		\frac{2 - 2 f(z) + z f'(z)}{z^2 f(z)} + \frac{1}{f(z)^2} A_{v}(z)^2
	\right) \sigma(z) = 0,
    \label{eq:eom_stationary_1}\\
	&A_{t}''(z) - \frac{2}{f(z)} A_{t}(z) \sigma(z)^2 = 0.
    \label{eq:eom_stationary_2}
\end{align}
In addition, we obtain the constraint equation
\begin{equation}
	\partial_{z} \theta(z) = - \frac{1}{f(z)} A_{t}(z).
\end{equation}
From this constraint, $A_{t}$ must vanish at $z = z_{h}$ for the regularity of the phase.
Imposing the vanishing Dirichlet condition on the scalar field, $\sigma(0)=0$, we can identify the condensate as $\abs{\expval{O_{2}}} = \sigma'(0)$.
The charge density is given by $\rho = - A_{t}'(0)$.

The solution associated with the symmetry restored phase is given by $\sigma(z) = 0$ with
\begin{equation}
    A_{t}(z) = \rho (z_{h} - z).
    \label{eq:At_solution}
\end{equation}
To obtain non-trivial solutions associated with the symmetry broken phase, we need to solve Eqs.~(\ref{eq:eom_stationary_1}) and (\ref{eq:eom_stationary_2}) numerically.
Figure \ref{fig:O2-rho} shows the order parameter as a function of the charge density for the stationary solutions.
Note that the system admits multiple solutions with $\sigma \neq 0$ associated with higher excited states \cite{Wang:2019caf, Wang:2019vaq}.
At fixed $T$, the curve having the largest order parameter corresponds to the ground state.
In our main analysis, we consider only the relaxation process under fixed $\rho < \rho_{\rm c}$, where the excited states do not exist, so these excited states have no influence.

\subsection{Quasinormal modes}
In order to analyze the quasinormal modes (QNMs) around the trivial solution given by $\psi =0$, we consider the linear perturbation $\psi \to \var{\psi}$.
In Fourier space, we obtain the linearized equation of motion as
\begin{equation}
\begin{aligned}
    \frac{
        i z^2 A_{t}'(z)+z f'(z)-2 f(z) + 2
        %- k^2 z^2
    }{z^2 f(z)}\var{\psi}(z)
    &+\\
    \frac{2 i A_{t}(z) +f'(z)+2 i \omega }{f(z)}
    \var{\psi}'(z) 
    &+\var{\psi}''(z) = 0.
\end{aligned}
\end{equation}
The background solution for $A_{t}$ is given by Eq.~(\ref{eq:At_solution}).
The equation for $\var{\psi}^{*}$ is obtained in the same manner.
We compute the QNMs by imposing $\var{\psi}=0$ at $z=0$ and the regularity condition at $z=z_{h}$.
%Note that the linear response is given by
%\[
%    \var{\expval{O_{2}}}
%    = \lim_{z\to 0} \var{\psi}_{,z} + i \mu \var{\psi} + i\omega \var{\psi}.
%\]
%The latter terms are required due to the differences of the coordinate and the gauge condition from the usual Poincare coordinate.
%The locations of the lowest QNM frequencies are shown in Fig.~\ref{fig:QNM_map}.
In Fig.~\ref{fig:QNM_profiles}, we show the radial profiles and the frequencies of the first four QNMs for $\rho = 3.0$.
The QNM frequencies of $\var{\psi}$ are not symmetric under $\omega \to -\omega^{*}$ but they correspond to the QNM frequencies of $\var{\psi}^{*}$.
The results are obtained using the Chebyshev pseudospectral method with $N_{z} = 50$.
Generally, in holography, QNM frequencies with large imaginary parts are difficult to calculate accurately.
%We have checked that the results in Fig.~\ref{fig:QNM_map} are robust even if we vary $N_{z}$.
We have checked that the results in Fig.~\ref{fig:QNM_profiles} are robust even if we vary $N_{z}$.

% \begin{figure}
% \centering
% \includegraphics[width=8cm]{./fig/QNMs.pdf}
% \caption{%
% 	Locations of the QNM frequencies in the complex plane $\omega = \omega_{\rm R} + i \omega_{\rm I}$ for $\rho_{\rm f} = 3.0$.
% }
% \label{fig:QNM_map}
% \end{figure}

\end{document}